\documentclass[twocolumn,amsmath,amssymb]{revtex4}

\usepackage{graphicx}
\usepackage{dcolumn}
\usepackage{bm}

\begin{document}

\preprint{APS/123-QED}

\title{Nonparametric model reconstruction for stochastic differential equation from discretely observed time-series data}
\author{Jun Ohkubo}
\email[Email address: ]{ohkubo@i.kyoto-u.ac.jp}
\affiliation{
Graduate School of Informatics, Kyoto University,\\
Yoshida Hon-machi, Sakyo-ku, Kyoto-shi, Kyoto 606-8501, Japan
}

\begin{abstract}
A scheme is developed for estimating state-dependent drift and diffusion coefficients 
in a stochastic differential equation from time-series data.
The scheme does not require to specify parametric forms for the drift and diffusion coefficients in advance.
In order to perform the nonparametric estimation,
a maximum likelihood method is combined with a concept based on a kernel density estimation.
In order to deal with discrete observation or sparsity of the time-series data,
a local linearization method is employed, which enables a fast estimation.
\end{abstract}

\maketitle

\section{Introduction}

Recently, it has been clarified that
stochastic nature in small systems such as cells
plays an important role in dynamics and behavior of biological systems
\cite{Rao2002,Elowitz2002,Kaern2005}.
In addition, due to recent experimental developments such as single-molecule spectroscopy,
it becomes possible to obtain a time-series data for various stochastic phenomena.
From a theoretical point of view, it is important to develop methods for analysis of the time-series data,
and actually there are many studies for the analysis of the single-molecule time-series
(for example, see Ref. \cite{Baba2007,Li2009,Miyazaki2011}).

Parameter estimations from observed time-series data are also important research topics.
If one obtains an experimental data for a specified biochemical system,
parameters in the specified biochemical system can be estimated from the experimental data.
The biochemical system could be modelled by using a master equation 
or a stochastic differential equation (a Langevin equation).
The master equation or the stochastic differential equation for the specified biochemical system
has some parameters (e.g., reaction rates).
If the reaction rates are estimated from the experimental data,
we will obtain a reconstructed model, which would reproduce the experimental data adequately.
The reconstructed model enables us to perform more detailed numerical simulations
and to have deep insights for the phenomenon.
In recent years, a discrete property or sparsity in observations has attracted much attention;
it would be difficult to completely observe the phenomenon and to obtain detailed time-series data,
and then we would perform the estimation from discretely observed time-series data.
For example, estimation procedures based 
on Markov Chain Monte Carlo methods \cite{Golightly2005,Boys2008,Wang2010}
and variational methods \cite{Ruttor2009,Vrettas2010} have been proposed
for the problem of the discrete observations.
(In addition, there is a recent review article \cite{Friedrich2011} 
for the estimation problem.)

Here, we consider the following situation:
We know that a time-series data can be modelled with a stochastic differential equation,
but specific forms of drift and diffusion coefficients of the stochastic differential equation
are unknown in advance.
That is, there is no prior knowledge about a time-series data,
except for some basic properties such as a memoryless property.
While there are some works for the parametric estimation based on 
a maximum likelihood method \cite{Kleinhans2005,Kleinhans2007},
simple applications of these parametric estimations are not suitable for our problem here, 
and a nonparametric estimation scheme is needed.
For example, we here focus on a bimodal distribution of a chemical substance.
Relations between the biomodal distributions and stochasticity have been discussed 
experimentally \cite{Acar2005,Toyooka2008} and theoretically \cite{Artyomov2007,Bishop2010}.
Although one may consider that the bimodal distribution is produced from a double-well potential system,
it has been known that the bimodal distribution can also be produced 
from a state-dependent noise \cite{Togashi2001,Ohkubo2008}.
In addition, a recent study indicates that such noise-induced bimodality may play an important role
in a decision making in a noisy environment \cite{Kobayashi2010,Kobayashi2011}.
In these situations, it is necessary to judge whether a bimodal distribution
is produced from a double-well potential system or a state-dependent noise,
and it is enough to estimate how the drift and diffusion coefficients 
of the stochastic differential equation depend on the state.
For the nonparametric estimations,
there are many studies in various research fields.
For example, in Ref.~\cite{Honisch2011},
a method based on estimations of Kramers-Moyal coefficients has been proposed.
However, in the Kramers-Moyal coefficients estimation,
adjoint Fokker-Planck equations should be solved numerically,
which needs additional computational costs.
In order to perform nonparametric estimations for complicated systems,
fast algorithms are required.

In the present paper, 
we develop a nonparametric model-reconstruction method for a stochastic differential equation
from a discretely observed time-series data.
A kind of local estimations is employed in order to extract the state-dependency 
in the nonparametric estimation.
In order to perform the local estimations efficiently, 
we propose a maximum likelihood method combined with a concept based on a kernel density estimation,
which has been studied a lot and widely used for nonparametric density estimation \cite{Silverman1998}.
The difficulty caused from the discrete observations 
is dealt with a local linearization method \cite{Biscay1996,Shoji1998},
which enables us to approximate a nonlinear stochastic differential equation
by a locally linear stochastic differential equation.
In addition, a useful `second-order' form suitable for the local linearization method is proposed.
We demonstrate that the combination of these ideas enables us to estimate
the state-dependent drift and diffusion coefficients 
from only a small set of discretely observed time-series data.

The present paper is constructed as follows.
In Sec.~II, we give problem settings and an example of a time-series data.
In Sec.~III, we briefly review a kernel density estimation,
and the scheme is reformulated from a different point of view;
we show that a maximum likelihood method reproduces the kernel density estimation adequately.
Section IV is the main part in the present paper;
an explicit estimation scheme based on the local linearization method is explained.
Examples of estimation results for the problem introduced in Sec.~II are given
in Sec.~V.
Section VI is the conclusion.

\section{Problem settings}

\begin{figure}
\begin{center}
  \includegraphics[width=80mm,keepaspectratio,clip]{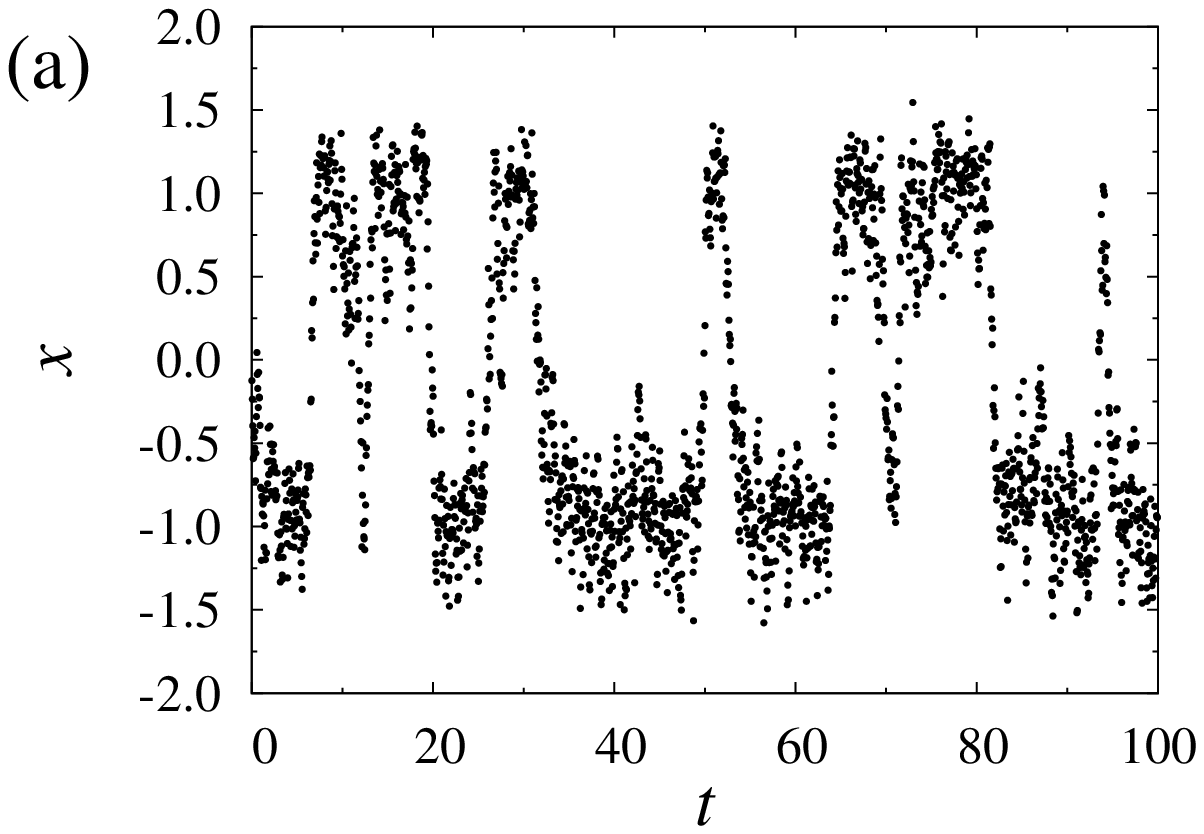}\\
  \includegraphics[width=80mm,keepaspectratio,clip]{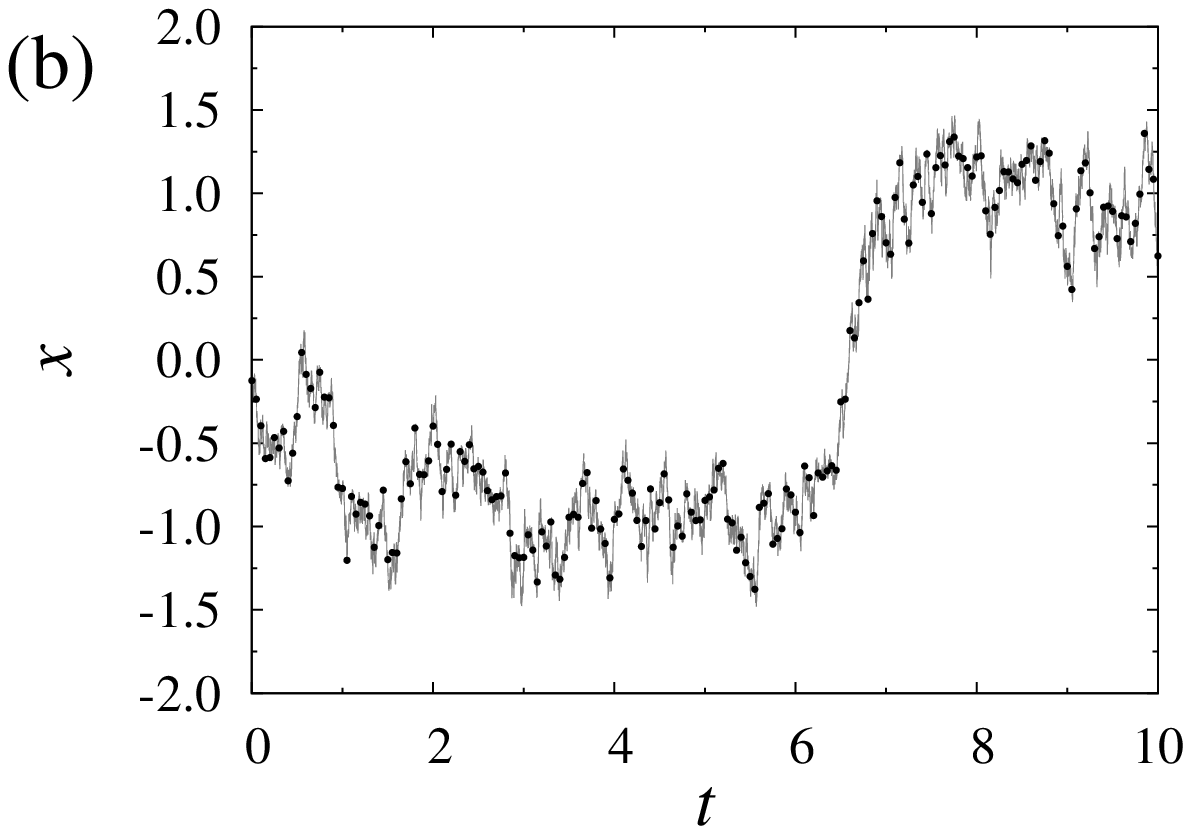}
\caption{
(a) Observed time-series data. There are totally $2000$ points observed discretely.
(b) Enlarged figure of (a). Each circle is an observation with $\Delta t = 0.05$,
and the solid thin line is an original path which is produced
using the Euler-Maruyama approximation with $\Delta t = 0.001$.
}
\label{fig_sample_path}
\end{center}
\end{figure}

The aim of our estimation problem is to reconstruct a stochastic differential equation
only from observed time-series data.
In order to demonstrate the estimation, we here use the following Ito-type stochastic differential equation
as a toy model:
\begin{align}
dx_t = f^\mathrm{true}(x_t)dt + g^\mathrm{true}(x_t) dW_t, 
\label{eq_original_sde_1}
\end{align}
where $W_t$ is a Wiener process, 
$f^\mathrm{true}(x)$ and $g^\mathrm{true}(x)$ are 
state-dependent drift and state-dependent diffusion coefficients
defined as
\begin{align}
f^\mathrm{true}(x) = -4 x^3 + 4 x, \quad g^\mathrm{true}(x) = 0.2 \sin( \pi x ),
\label{eq_original_sde_2}
\end{align}
respectively.
This model has state-dependent drift and diffusion coefficients.
In addition, this form of the stochastic differential equation suggests that
the potential landscape is a double-well form, and there are two stable points
around $x = 1$ and $x = -1$.
Using the Euler-Maruyama scheme \cite{Kloeden}, we generate an `original' path
for the stochastic differential equation \eqref{eq_original_sde_1}.
In the generation of the original path, we employ a time interval of $\Delta t = 0.001$.
After the generation of the original path,
we sampled data points discretely, which corresponds to discrete observations.
Although the time interval of the observation can be varied,
we here take data points at equally spaced time interval $\Delta t = 0.05$ for simplicity.
The generated time-series data are depicted in Fig.~\ref{fig_sample_path}.
Hereafter, we must forget the stochastic differential equation \eqref{eq_original_sde_1}
and the state-dependency of the drift and diffusion coefficients in Eq.~\eqref{eq_original_sde_2},
and only the time-series data in Fig.~\ref{fig_sample_path} is focused and analyzed.

Our aim is to analyze the time-series data in Fig.~\ref{fig_sample_path}
under a situation that we do not have any information about the original model.
A first guess may be as follows:
It seems that there are two stable points around $x \sim 1.0$ and $x \sim -1.0$,
and the time spent around $x \sim -1.0$ seems to be a little longer than that around $x \sim 1.0$.
Is the difference caused by a potential landscape, or other reasons?
A simple way to solve this problem is 
to reconstruct an explicit model which reproduces the time-series data.
Hence, the problem settings are as follows:
Estimate $\hat{f}(x)$ and $\hat{g}(x)$ in
\begin{align}
dx_t = \hat{f}(x_t)dt + \hat{g}(x_t) dW_t
\label{eq_sde_to_be_estimated}
\end{align}
from a time-series data $\{ (X_i, T_i) | i = 1, \dots, N\}$,
where each data point $X_i$ is observed at time $T_i$,
and $N$ is the total number of the observed data points.
Note that there is no prior knowledge about the state-dependencies of $\hat{f}(x)$ and $\hat{g}(x)$,
except that these coefficients are time-independent.
It could be possible to deal with time-dependent cases,
but it is beyond the scope of the present paper.


\section{Kernel density estimation}

In order to develop a scheme to estimate the drift and diffusion coefficients
in Eq.~\eqref{eq_sde_to_be_estimated},
we firstly explain a kernel density estimation,
which gives us many insights for our final aim.

\subsection{Brief review}

A first and simple step to analyze a time-series data is to construct
a probability density $p(x)$ for the observed data.
A histogram, in which the $x$ coordinate is split into several bins and the numbers of data points
within the bins are counted,
is a simple method to estimate the probability density $p(x)$.
However, the histogram is based on a discrete approximation.
One of the most widely used method for nonparametric density estimation
is a kernel density estimation \cite{Rosenblatt1956,Silverman1998}.
The kernel density estimation has been recently studied in the context of biophysics
\cite{Bura2009}, in which applications for the forced unfolding
and unbinding data for proteins are discussed.

In the kernel density estimation,
a non-negative real function $K(x)$, i.e., a kernel function, is used.
The kernel function satisfies 
the normalization condition, $\int_{-\infty}^{\infty}K(x) dx = 1$,
and it has a zero first moment, $\int_{-\infty}^{\infty} x K(x) dx = 0$,
and a finite second moment $\int_{-\infty}^{\infty} x^2 K(x) dx < \infty$.
In the present paper, a parameter $h$, a so-called bandwidth, 
is introduced explicitly in the kernel function,
and we write the kernel function as $K_h(x) \equiv K(x/h)/h$.
Using the kernel function,
the probability density is estimated as
\begin{align}
\hat{p}(x) = \frac{1}{N} \sum_{i=1}^N K_h(x-X_i).
\label{eq_density_estimator}
\end{align}
There are some kernel functions,
and in the present paper, we consider only a Gaussian kernel
$K_h(x) = (1/\sqrt{2\pi} h) \exp(-x^2 / (2 h^2))$,
which has been widely used.

The remaining task for the kernel density estimation
is the choice of the bandwidth $h$ of the kernel function.
Various choices have been studied,
and a famous data-driven method is a method based on a cross-validation \cite{Rudemo1982}.
In the cross-validation method,
the following risk function is minimized with respect to the bandwidth $h$ for the Gaussian kernel:
\begin{align}
\hat{Q} = A + B \sum_{i < j} \left[
\exp\left( - \Delta_{ij}^2/4 \right) 
- C \exp\left( - \Delta_{ij}^2/2\right)
\right],
\label{eq_risk_function}
\end{align}
where
\begin{align*}
&A = (2 N h \sqrt{\pi})^{-1}, \quad B = (N^2 h \sqrt{\pi})^{-1}, \\
&C = 2\sqrt{2} N / (N-1), \quad \Delta_{ij} = (X_i-X_j)/h.
\end{align*}

\begin{figure}
\begin{center}
  \includegraphics[width=80mm,keepaspectratio,clip]{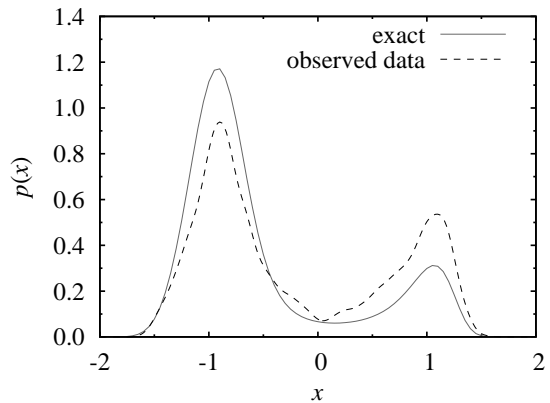}
\caption{
Probability density function $p(x)$. The solid line corresponds to an exact solution of
Eq.~\eqref{eq_original_sde_1}.
The estimated density from the observed time-series data
is depicted with the dashed line.
The bimodality of the density is reproduced,
but heights of the peaks are different from the exact one
because of the small size of the data in Fig.~\ref{fig_sample_path}.
}
\label{fig_sample_distribution}
\end{center}
\end{figure}

Figure~\ref{fig_sample_distribution} shows the estimated probability density.
Here, we used $h = 0.06476$, which is selected based on the risk function \eqref{eq_risk_function}.
The solid line corresponds to an exact solution obtained from Eq.~\eqref{eq_original_sde_1}.
Because of the small size of the data in Fig.~\ref{fig_sample_path},
there are differences between the estimated probability density and the exact solution.
We note that if there is a longer time-series data, better estimate results are obtained.

\subsection{Kernel density estimation through maximum likelihood estimation}

The estimation of the probability density is not the aim in the present paper,
but it is helpful for us to reformulate the kernel density estimation 
from the view point of a maximum likelihood estimation;
this discussion gives us a way to reconstruct a stochastic differential equation
without any prior knowledge.

The maximum likelihood estimation enables us to estimate parameters in a statistical model
(for example, see Ref.~\cite{Hogg}).
In the context of the density estimation,
the probability density plays a role as the parameters,
and we seek `probable' probability density function $\hat{p}(x)$ from observed data.

Due to the concept of the kernel function, a contribution from one observed data point
should be distributed according to the kernel function.
For example, assume that we observe a data point $X = 1.0$.
Although we have only one data point, here we introduce many `replicas' for the observation.
Each replica has a different `pseudo-observation'.
For instance, when there are four replicas and the replicas have pseudo-data-points
$X = 1.0, 1.2, 0.7, 1.2$,
the probability with which we observe the total replicas is
\begin{align*}
p(X=1.0) ( p(X=1.2) )^2 p(X=0.7).
\end{align*}
There are four pseudo-observations for only one real observation, 
and then one may consider the fourth root,
\begin{align*}
(p(X=1.0) )^{1/4} ( p(X=1.2) )^{2/4}  (p(X=0.7))^{1/4}.
\end{align*}
Note that the frequency of `pseudo-observation' of $X$ is the power index of each probability.
Extending this discussion and 
using the kernel function instead of the frequency of `pseudo-observation',
we construct the `kernelized' likelihood function, as follows.
Using a discretization of the $x$ coordinate as $x = j \Delta x$,
a probability with which we observe a `distributed' data point $X_i$ is written as
\begin{align*}
\lim_{\Delta x \to 0} \prod_{j = -\infty}^{\infty} p(j \Delta x)^{K_h(j \Delta x - X_i)},
\end{align*}
Note that if we use a delta-like function as the kernel function,
only a contribution from $p(X_i)$ remains, and an usual interpretation
without the kernel function is recovered.
Using a notation $\bm{X} = \{X_1, \dots, X_N\}$,
a likelihood function $L(\hat{p}(x)| \bm{X})$ is written as
\begin{align}
L(\hat{p}(x)| \bm{X}) = 
\lim_{\Delta x \to 0} \prod_{i=1}^N \prod_{j_i = -\infty}^{\infty} \hat{p}(j_i \Delta x)^{K_h(j_i \Delta x - X_i)},
\end{align}
where the {\it hat} of $\hat{p}$ means that this is just the parameter to be estimated.
Hence, the log-likelihood function is
\begin{align}
l(\hat{p}(x) | \bm{X}) = 
\sum_{i=1}^N \int_{-\infty}^{\infty} dx_{i} \left[ \log \hat{p}(x_i) \right] K_h(x_i-X_i).
\label{eq_log_l_density}
\end{align}

In order to obtain the maximum likelihood estimates for $\hat{p}(x)$,
we take a functional derivative of Eq.~\eqref{eq_log_l_density} with respect to $\hat{p}(x)$
under a constraint $\int dx \hat{p}(x) = 1$.
A Lagrange multiplier $\lambda$ is introduced and
we consider a maximization of the following function:
\begin{align*}
\sum_{i=1}^N \int_{-\infty}^{\infty} dx_{i} \left[ \log \hat{p}(x_i) \right] K_h(x_i-X_i) 
+ \lambda \left( \int dx \hat{p}(x) - 1 \right).
\end{align*}
Taking the functional derivative with respect to $\hat{p}(x)$ and setting
the functional derivative is equal to zero, we obtain 
\begin{align}
\sum_{i=1}^N \frac{1}{\hat{p}(x)} K_h(x-X_i) + \lambda = 0,
\end{align}
and therefore
\begin{align}
\hat{p}(x) = \frac{1}{\lambda} \sum_{i=1}^N K_h(x-X_i).
\label{eq_log_l_tmp}
\end{align}
Inserting Eq.~\eqref{eq_log_l_tmp} into the constraint condition,
we have
\begin{align}
1 = \frac{1}{\lambda} \int_{-\infty}^{\infty} dx \sum_{i=1}^N K_h(x-X_i) = \frac{N}{\lambda}.
\end{align}
Hence, $\lambda = N$ and we recover Eq.~\eqref{eq_density_estimator}.

The above discussion indicates that
the concept of `distributed' data points in `pseudo-observation' corresponds to the kernel function.
It would be expected that 
the combination of the `distributed' data points
and the maximum likelihood method gives us more flexible estimation scheme
than the usual one.

It is straightforward to extend the above discussions to an estimation of a conditional density.
Given $\mathcal{D} = \{(X_1,Y_1),\dots,(X_N,Y_N) \}$, 
a sample of independent observations from the distribution of $(X,Y)$,
we want to obtain the estimation of the conditional density $\hat{p}(y|x)$.
In this case, the log-likelihood function should be set as
\begin{align}
&l(\hat{p}(y|x) | \mathcal{D}) \nonumber \\
&= 
\sum_{i=1}^N \int_{-\infty}^{\infty} dy_i
\left[ \log \hat{p}(y_i|x) \right] \tilde{K}_w(y_i-Y_i) K_W(x-X_i),
\end{align}
where $\tilde{K}_w(y)$ is a kernel function for the $y$ coordinate, and
$K_W(x)$ is that for the $x$ coordinate.
Using the similar discussion as Eq.~\eqref{eq_log_l_density},
we obtain the following conditional density estimator
\begin{align}
\hat{p}(y|x) = \frac{\sum_{i=1} K_W(x-X_i) \tilde{K}_w(y-Y_i)}{\sum_{i=1}^{N} K_W (x-X_i)},
\end{align}
which is the same as a conventional conditional density estimator
\cite{Hyndman1996,Bashtannyk2001}.

\begin{figure}
\begin{center}
  \includegraphics[width=80mm,keepaspectratio,clip]{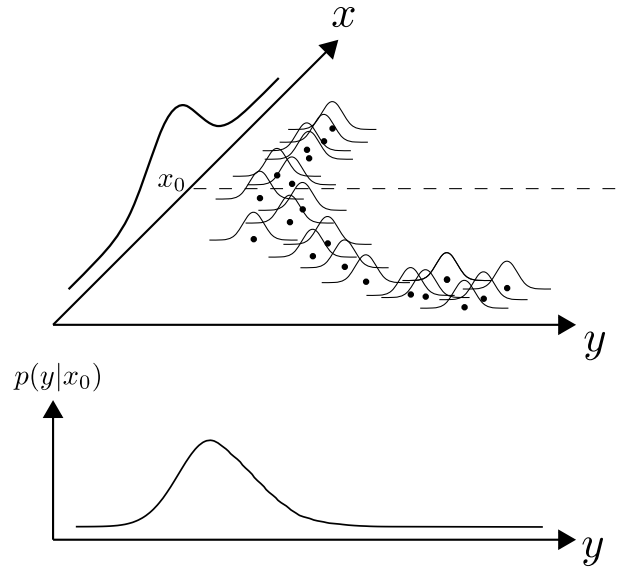}
\caption{
A schematic illustration of the kernel density estimate $p(y|x_0)$.
Each data point has a weight which is smoothly distributed according to the kernel function.
The conditioning on $x=x_0$ is carried out by another kernel function in the $x$ coordinate.
}
\label{fig_explanation_kernel}
\end{center}
\end{figure}

Figure~\ref{fig_explanation_kernel} shows a schematic illustration of the kernel
conditional density estimator.
We here consider a conditional density $p(y|x_0)$.
The kernel functions $\tilde{K}_w$ with bandwidth $w$ are distributed according to the observations.
The conditioning $x=x_0$ is carried out by another kernel function in the $x$ coordinate.
The kernel function $K_W$ has a bandwidth $W$, and the center of the kernel function
is at $x=x_0$.
The estimation of the conditional density is performed
by summing the $N$ kernel functions in the $y$ coordinate, weighted by the kernel function
in the $x$ coordinate.

For $\{(X_1,Y_1), \dots, (X_N,Y_N)\}$ being random samples
from a population having a density $p(x,y)$,
there are many studies for estimating conditional densities
and properties of the conditional densities \cite{Fan1992,Ruppert1994,Fan1996}.
The problem in the present paper is different from these studies;
$\{(X_1,Y_1), \dots, (X_N,Y_N)\}$ are not generated from an identical density; 
see the next section.
The maximum likelihood method developed in this section
enables us to deal with the nonparametric estimation
for a stochastic differential equation,
and we will propose the method in the next section.

\section{Proposed method}

\subsection{Basics}

We are now ready for constructing a method to estimate drift and diffusion coefficients
in a stochastic differential equation from a time-series data.

Firstly, the observed data $\{(X_i,T_i) | i = 1, \dots, N\}$
is converted to a slightly different form
$\mathcal{D} = \{ (X_i, \Delta X_i, \Delta t_i) | i = 1, \dots, N-1\}$,
where $\Delta X_i = X_{i+1} - X_i$ and $\Delta t_i = T_{i+1} - T_i$.
This conversion means that 
when a current coordinate is $X_i$, 
we have an amount of change $\Delta X_i$ during the time interval $\Delta t_i$.
For each $i$, the time interval can be varied in general,
and then the amount of change $\Delta X_i$ depends on $\Delta t_i$;
$\{(X_i, \Delta X_i)\}$ is not from an identical density.
Due to an assumption that the time-series has a Markov property,
$\{(X_i, \Delta X_i)\}$ are independent each other.

Secondly, we consider a conditional probability density
$p(\Delta x| x, \bm{\theta}_x)$,
which has a set of parameters $\bm{\theta}_x$.
The parameters $\bm{\theta}_x$ depends on a specific coordinate $x$,
and the dependency of $\bm{\theta}_x$ on $x$ is unknown in advance.
Note that the parametrization of the conditional probability density
is not related to the parametrization
of drift and diffusion coefficients in the stochastic differential equation
in Sec. II.
Using the localized parameters $\bm{\theta}_x$,
the drift and diffusion coefficients only for a specific coordinate $x$ could be estimated.

Thirdly, we consider the following log-likelihood function:
\begin{align}
l(\bm{\theta}_x | \mathcal{D})
=&\sum_{i=1}^{N-1} \int_{-\infty}^{\infty} d(\Delta x_i)
\left[ \log p(\Delta x_i| x, \bm{\theta}_x) \right]  \nonumber \\
&\quad \times \tilde{K}_w(\Delta x_i - \Delta X_i) K_W(x-X_i).
\end{align}
Maximizing the above log-likelihood function with respect to
the parameters $\bm{\theta}_x$,
it is possible to estimate the localized parameter $\bm{\theta}_x$
adequately.

A remaining task is to specify the
conditional probability density $p(\Delta x| x, \bm{\theta}_x)$.
If a stochastic differential equation is linear,
the conditional probability density is expressed
as a normal distribution,
i.e., $\Delta X_i$ obeys
a normal distribution with mean $E_i$ and variance $V_i$,
where $E_i$ and $V_i$ depend on $X_i$ and $\Delta t_i$.
For a nonlinear stochastic differential equation,
the description based on the normal distribution is impossible in general.
However, if the conditional probability density cannot be written
as the normal distribution, 
the calculation scheme would become very complicated.
Hence, we here assume that
the conditional probability density is written as the normal distribution.
The restriction with the normal distribution seems to be severe,
but we will discuss a method to approximate the nonlinear stochastic differential equation
to a `locally' linear stochastic differential equation.
The assumption of the Gaussian form
gives the conditional probability density,
\begin{align}
\log p(\Delta x_i| x, \bm{\theta}_x)
= - \frac{1}{2}
\left\{
\frac{(\Delta x_i - E_i)^2}{V_i} + \log (2\pi V_i)
\right\},
\end{align}
where $E_i$ and $V_i$ depend
on $X_i$, $\Delta t_i$, and the parameters $\bm{\theta}_x$.

\subsection{Simple estimation}

We here discuss the most simple case,
i.e., $E_i = \mu_{x_0} \Delta t_i$, $V_i = \sigma_{x_0}^2 \Delta t_i$,
and $\bm{\theta}_{x_0} = \{\mu_{x_0}, \sigma_{x_0}\}$.
This means that
for the estimation at $x = x_0$, we assume
a stochastic differential equation 
with constant drift and diffusion coefficients.
Note that this constant property is assumed only for the estimation at $x=x_0$,
and we do not assume that $\hat{f}(x)$ and $\hat{g}(x)$ in Eq.~\eqref{eq_sde_to_be_estimated}
are constant for all $x$.
Repeating the estimation of $\mu_{x_0}$ and $\sigma_{x_0}$ 
for various point $x=x_0$, 
we obtain the estimated drift and diffusion coefficients as
$\hat{f}(x) = \mu_{x}$ and $\hat{g}(x) = \sigma_x$, respectively.

The above procedure is enough for the estimation,
but here we give a discussion for the choice of the kernel bandwidth.
For simplicity, we here assume that
the kernel $\tilde{K}_w$ is the Gaussian kernel,
and that $\Delta t_i = 1$ for all $i$.
In addition, we consider a simple case in which $X_i = x_0$ for all $i$,
i.e., all current coordinates are at $x_0$.
Hence, the kernel function $K_W$ is constant for all $i$.
In this case,
the following log-likelihood function is obtained:
\begin{align}
&l(\bm{\theta}_{x_0} | \mathcal{D}) \nonumber \\
&\propto - \frac{1}{2} \sum_{i=1}^{N-1} 
 \int_{-\infty}^{\infty} d(\Delta x_i) \nonumber \\
& \times \left\{ 
\frac{(\Delta x_i - \mu_{x_0})^2}{\sigma_{x_0}^2} + \log( 2\pi \sigma_{x_0}^2 )
\right\} 
 \tilde{K}_w(\Delta x_i - \Delta X_i) \nonumber \\
&= - \frac{1}{2} \sum_{i=1}^{N-1} 
\left\{ 
\frac{(\Delta X_i - \mu_{x_0})^2}{\sigma_{x_0}^2} 
+ \frac{w^2}{\sigma_{x_0}^2} 
+ \log( 2\pi \sigma_{x_0}^2 )
\right\} 
\end{align}
The maximum likelihood method gives
\begin{align}
\mu_{x_0} = \frac{1}{N-1} \sum_{i=1}^{N-1} \Delta X_i
\end{align}
and
\begin{align}
\sigma_{x_0}^2 = \frac{1}{N-1} \sum_{i=1}^{N-1}(\Delta X_i - \mu_{x_0})^2 + w^2.
\label{eq_simple_sigma}
\end{align}
Here, note that the unbiased estimator for the variance is given as
\begin{align}
\bar{\sigma}_{x_0}^2 = \frac{1}{N-2} \sum_{i=1}^{N-1} (\Delta X_i - \mu_{x_0})^2.
\label{eq_unbiased_sigma}
\end{align}
Comparing Eq.~\eqref{eq_simple_sigma} with Eq.~\eqref{eq_unbiased_sigma},
it is clear that the bandwidth $w$ should be taken small enough for large $N$.
In our problem settings in the present paper,
it is possible to consider that $w \simeq 0$.
Hence, the kernel $\tilde{K}_w(\Delta x_i - \Delta X_i)$ 
is replaced as a Dirac delta function $\delta(\Delta x_i - \Delta X_i)$ hereafter.


Combining the above all discussions,
we obtain the following simple estimation scheme:
\begin{enumerate}
\item[] \textit{Algorithm 1 (Simple method)}
\item Set a bandwidth $W$.
\item For a point $x_0$, maximize the following log-likelihood function
with respect to $\bm{\theta}_{x_0} = \{\mu_{x_0}, \sigma_{x_0}\}$:
\begin{align}
&l(\bm{\theta}_{x_0}|\mathcal{D}) \nonumber \\
&=
- \frac{1}{2} \sum_{i=1}^{N-1} 
\left\{ 
\frac{(\Delta X_i - \mu_{x_0} \Delta t_i)^2}{ \sigma_{x_0}^2 \Delta t_i}
+ \log(2\pi \sigma_{x_0}^2)
\right\}\nonumber \\
& \qquad \times  K_W(x - X_i),
\end{align}
i.e., calculate the following quantities:
\begin{align}
\mu_{x_0} &= \frac{\sum_{i=1}^{N-1} \Delta X_i K_W(x-X_i)}{\sum_{i=1}^{N-1} \Delta t_i K_W(x-X_i)}, \\
\sigma_{x_0}^2 &= 
\frac{\sum_{i=1}^{N-1} (\Delta X_i - \mu_{x_0}\Delta t_i)^2 K_W(x-X_i) / \Delta t_i}
{\sum_{i=1}^{N-1} K_W(x-X_i)}.
\end{align}
\item Repeat 2 for various $x_0$.
\item Estimate the drift and diffusion coefficients as
$\hat{f}(x) = \mu_{x}$ and $\hat{g}(x) = \sigma_{x}$.
\end{enumerate}

\subsection{Local linearization method and `second-order' approximation}

In Sec.~IV~B, 
the \textit{local} drift and diffusion coefficients have simple forms,
so that we easily solve the stochastic differential equation explicitly.
Note that if we have nonlinear drift and diffusion coefficients,
we cannot obtain an explicit solution
for the stochastic differential equation exactly in general.
However, using a local linearization method \cite{Biscay1996,Shoji1998},
it is possible to obtain the approximate solution with a Gaussian form for the nonlinear 
stochastic differential equation.
Hence, the estimation scheme developed in Sec.~IV~A is available
even for the nonlinear cases.
We can assume arbitrary drift and diffusion coefficients,
and a `second-order' approximation, which will be introduced soon, is 
one of tractable schemes.

We firstly note that the diffusion coefficient in the stochastic differential
equation must be positive for all $x$;
this fact needs an additional constraint for the optimization procedures.
We, therefore, use the following `second-order' approximation;
$\hat{f}(x)$ and $\hat{g}(x)$ around $x_0$
is approximated as
\begin{align}
&\hat{f}_{x_0}(x) = \mu^{(0)}_{x_0} + \mu^{(1)}_{x_0}(x-x_0) + \frac{1}{2} \mu^{(2)}_{x_0} (x-x_0)^2, \\
&\hat{g}_{x_0}(x) = \exp\left( 
s^{(0)}_{x_0} + s^{(1)}_{x_0}(x-x_0) + \frac{1}{2} s^{(2)}_{x_0} (x-x_0)^2 
\right),
\label{eq_assumed_g}
\end{align}
and $\bm{\theta}_{x_0} = \{ \mu^{(0)}_{x_0}, \mu^{(1)}_{x_0}, \mu^{(2)}_{x_0}, 
s^{(0)}_{x_0}, s^{(1)}_{x_0}, s^{(2)}_{x_0} \}$.
Due to the exponential form in Eq.~\eqref{eq_assumed_g},
the diffusion coefficient $\hat{g}_{x_0}(x)$ is positive for all $x$.

We comment that 
a final estimation for the drift and diffusion coefficients
should be performed as $\hat{f}(x) = \mu^{(0)}_x$, 
and $\hat{g}(x) = \exp(s^{(0)}_{x})$, respectively,
as discussed in Sec.~IV~B.

A stochastic differential equation with the state-dependent drift coefficient $\hat{f}_{x_0}(x)$
and the state-dependent diffusion coefficient $\hat{g}_{x_0}(x)$
is nonlinear,
and the local linearization method gives
an analytical solution in a Gaussian form.
As a result, the conditional probability density $p(\Delta x_i| x, \bm{\theta}_x)$
can be written as a Gaussian distribution.
We briefly explain the local linearization method in the Appendix,
and only the consequence is shown here.
We note that the kernel function $\tilde{K}_w$ is replaced 
with the Dirac delta function according to the discussion in Sec.~IV~B.
The estimation scheme based on the local linearization method is as follows:
\begin{enumerate}
\item[] \textit{Algorithm 2 (LL method)}
\item Set a bandwidth $W$.
\item For a point $x_0$, maximize the following log-likelihood function
with respect to
$\bm{\theta}_{x_0} = \{ \mu^{(0)}_{x_0}, \mu^{(1)}_{x_0}, \mu^{(2)}_{x_0}, 
s^{(0)}_{x_0}, s^{(1)}_{x_0}, s^{(2)}_{x_0} \}$:
\begin{align}
&l(\bm{\theta}_{x_0}|\mathcal{D}) \nonumber \\
&=
- \frac{1}{2} \sum_{i=1}^{N-1} 
\left\{ 
\frac{ ( \phi_{x_0} - E_i )^2}{V_i} + \log(2\pi V_i)
\right.\nonumber \\
& \left. 
\qquad - s^{(0)}_{x_0} - s^{(1)}_{x_0}(X_i-x_0) 
- \frac{s^{(2)}_{x_0}}{2} (X_i-x_0)^2 \right\} \nonumber \\
& \qquad \times  K_W(x - X_i),
\label{eq_LL_log_likelihood}
\end{align}
where
\begin{align}
&\phi_{x_0}  \nonumber \\
&= \int_{X_i-x_0}^{X_i + \Delta X_i-x_0} du 
\exp \left( - s^{(0)}_{x_0} - s^{(1)}_{x_0}u - \frac{1}{2} s^{(2)}_{x_0} u^2 \right) ,
\end{align}
and 
\begin{align}
F_i =& \frac{\hat{f}_{x_0}(X_i)}{\hat{g}_{x_0}(X_i)} 
- \frac{\hat{g}_{x_0}(X_i)}{2} \left(s^{(1)}_{x_0} + s^{(2)}_{x_0} (X_i-x_0) \right), \\
L_i =& \mu^{(1)}_{x_0} + \mu^{(2)}_{x_0} (X_i-x_0) \nonumber \\
&- \hat{f}_{x_0}(X_i) \left(s^{(1)}_{x_0} + s^{(2)}_{x_0} (X_i-x_0) \right) \nonumber \\
&- \left(\hat{g}_{x_0}(X_i) \right)^2 \left( 
\frac{s^{(2)}_{x_0}}{2} + \frac{\left(s^{(1)}_{x_0} + s^{(2)}_{x_0} (X_i-x_0) \right)^2}{2}
\right),\\
M_i =& \frac{\hat{g}_{x_0}(X_i)}{2} \left[
\mu^{(2)}_{x_0} - s^{(2)}_{x_0} \hat{f}_{x_0}(X_i)   \right. \nonumber \\
&- \left(\mu^{(1)}_{x_0} + \mu^{(2)}_{x_0} (X_i-x_0) \right)
\left( s^{(1)}_{x_0} + s^{(2)}_{x_0} (X_i-x_0) \right) \nonumber \\
&- 2 \left(\hat{g}_{x_0} (X_i)\right)^2 
s^{(2)}_{x_0} \left(s^{(1)}_{x_0}+s^{(2)}_{x_0} (X_i-x_0) \right) \nonumber \\
&\left. - \left(\hat{g}_{x_0} (X_i)\right)^2 
\left(s^{(1)}_{x_0}+s^{(2)}_{x_0} (X_i-x_0) \right)^3
\right], \\
E_i =& \frac{F_i}{L_i} \left(e^{L_i \Delta t_i}-1 \right) 
+ \frac{M_i}{L_i^2}\left( e^{L_i \Delta t_i} - 1 - L_i \Delta t_i\right),  \\
V_i =& \frac{e^{2L_i \Delta t_i} - 1}{2 L_i}.
\end{align}
\item Repeat 2 for various $x_0$.
\item Estimate the drift and diffusion coefficients as
$\hat{f}(x) = \mu^{(0)}_{x}$ and $\hat{g}(x) = \exp( s^{(0)}_{x})$.
\end{enumerate}
For step $2$, various standard numerical maximization or minimization algorithms are available.
We note that the function $\phi_{x_0}$
can be written using an error function or an imaginary error function.

\section{Estimation results}

\begin{figure*}
\begin{center}
  \includegraphics[width=80mm,keepaspectratio,clip]{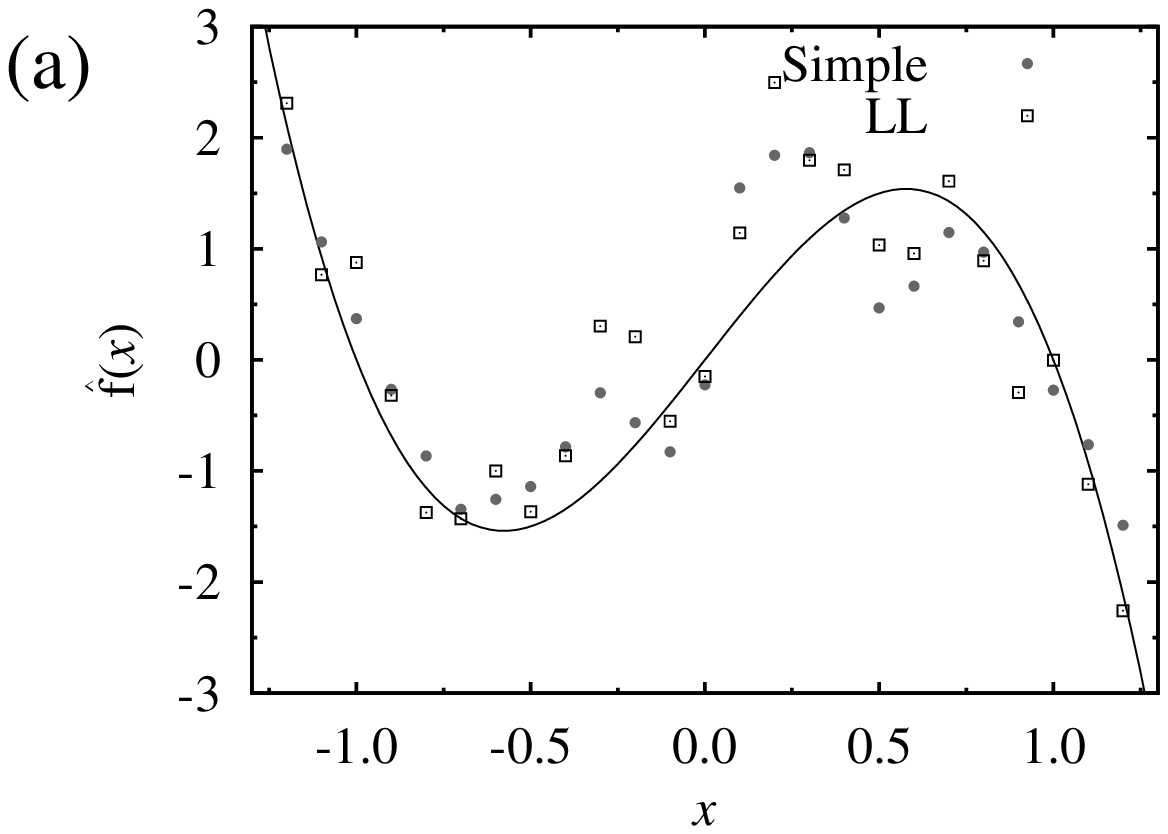}
  \includegraphics[width=80mm,keepaspectratio,clip]{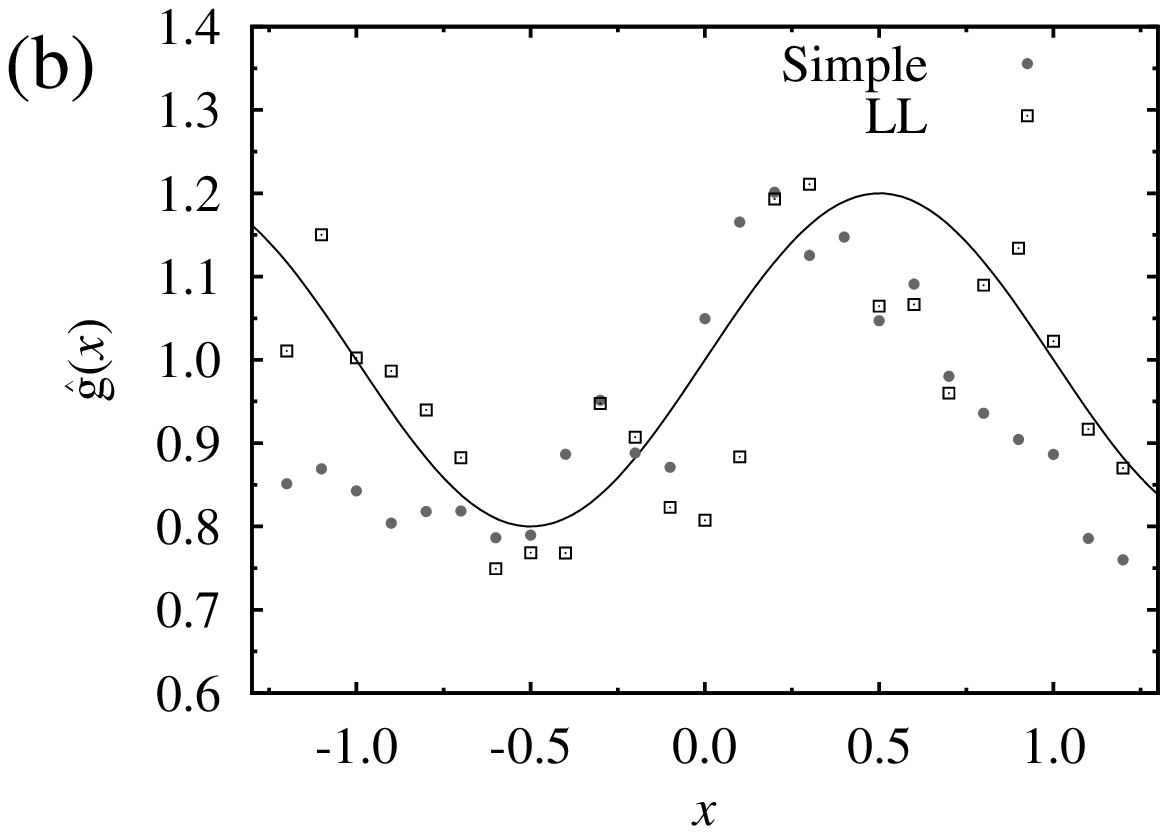}\\
  \includegraphics[width=80mm,keepaspectratio,clip]{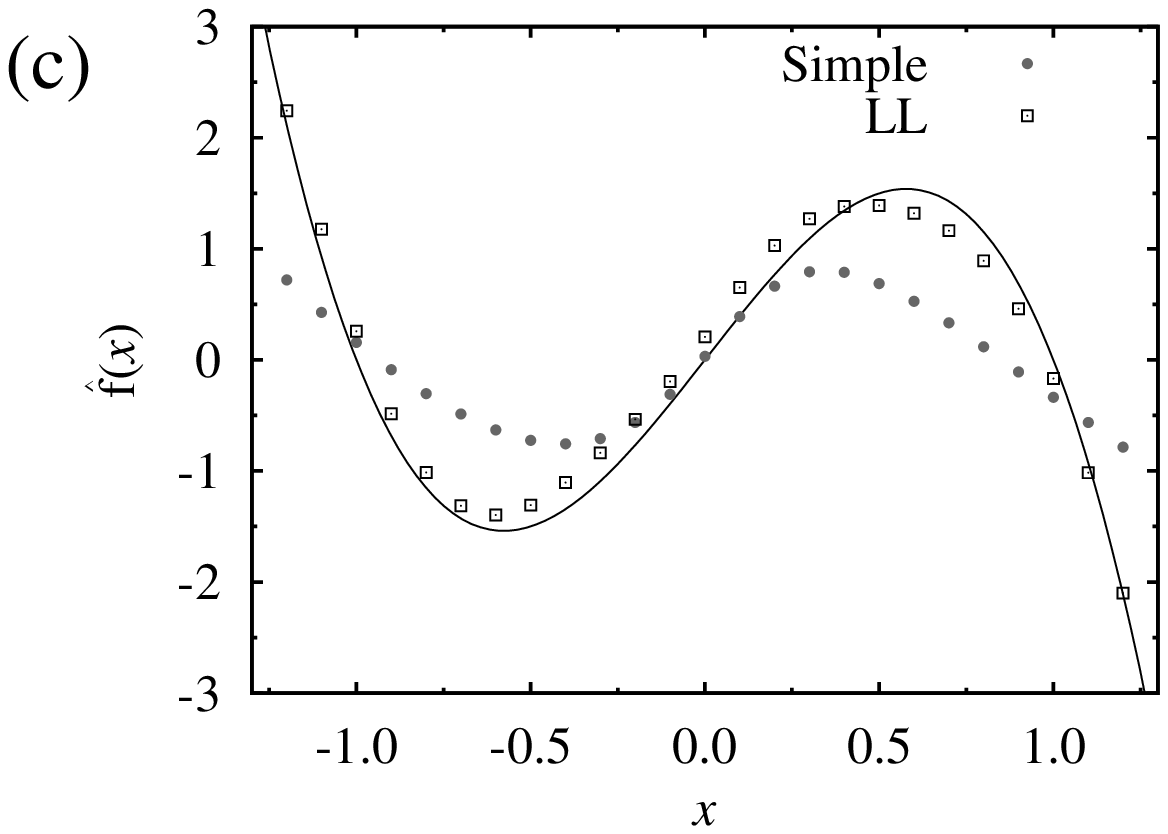}
  \includegraphics[width=80mm,keepaspectratio,clip]{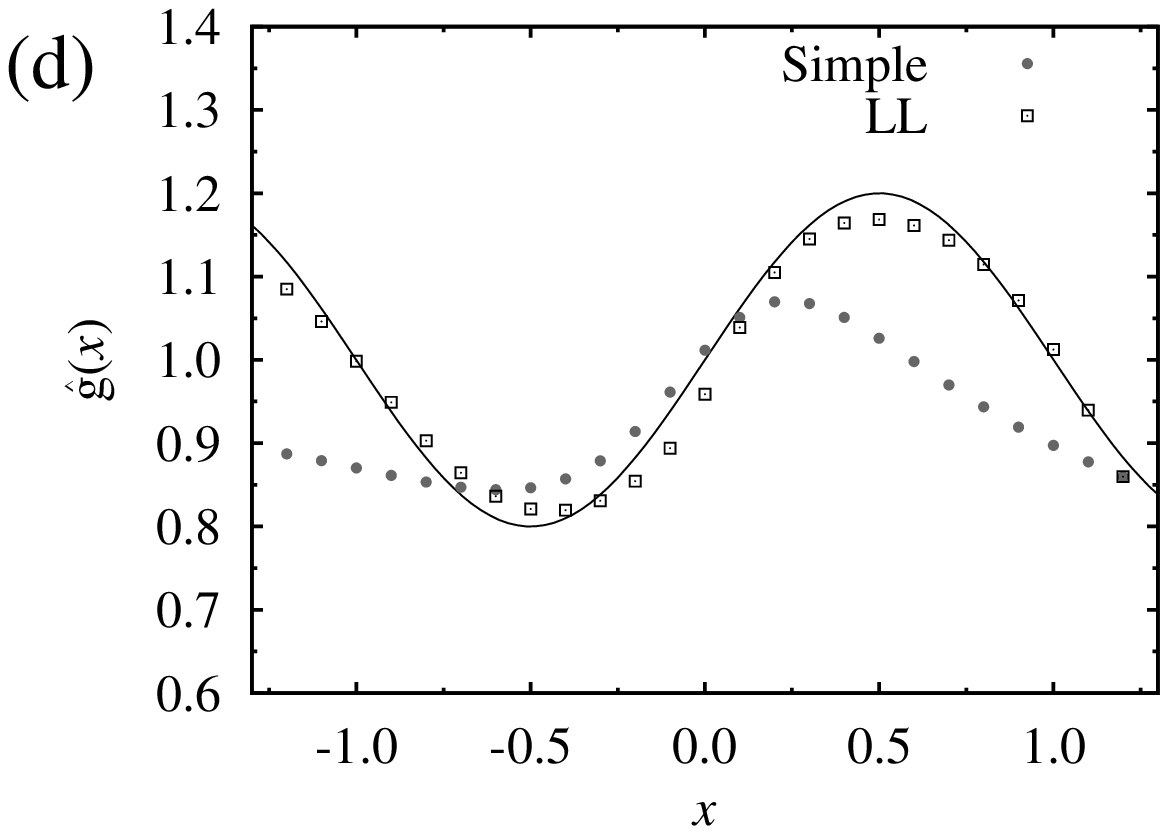}\\
  \includegraphics[width=80mm,keepaspectratio,clip]{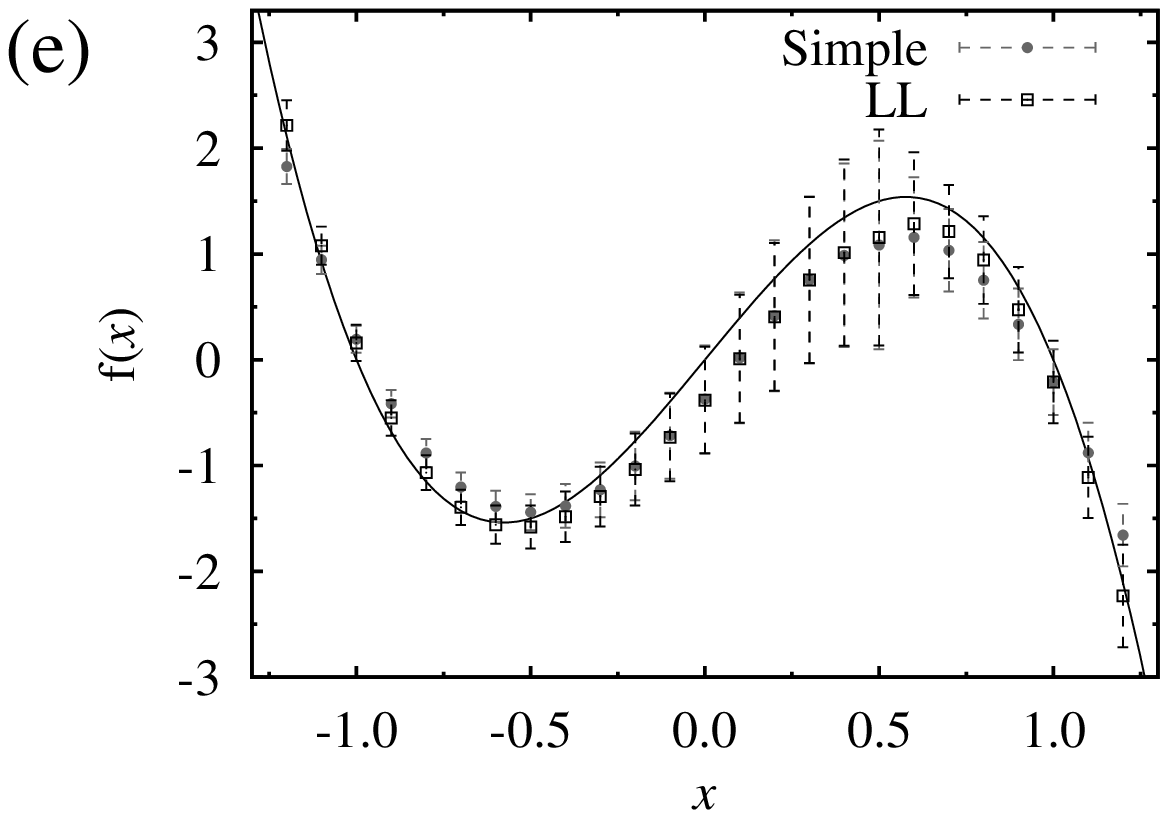}
  \includegraphics[width=80mm,keepaspectratio,clip]{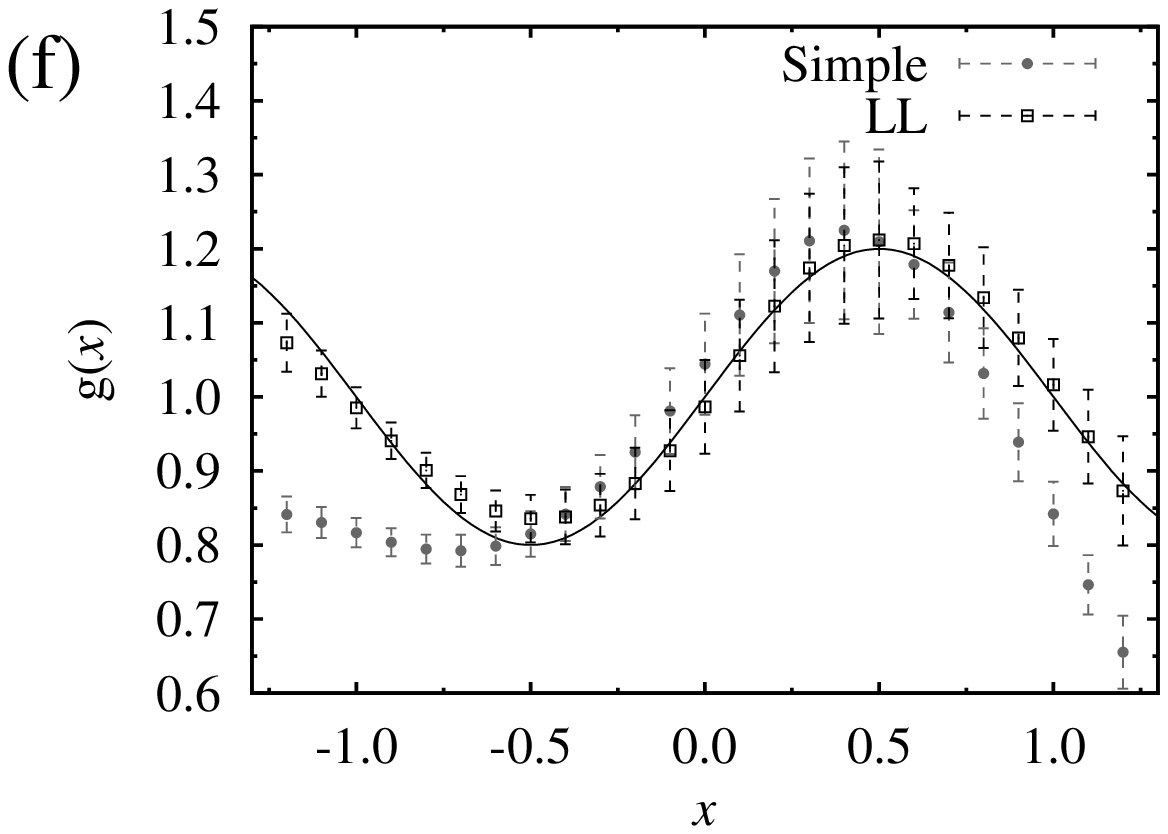}
\caption{
Estimation results 
for the drift coefficients $\hat{f}(x)$ [panels (a), (c), and (e)], 
and for the diffusion coefficients $\hat{g}(x)$ [panels (b), (d), and (f)].
Panels (a)-(d) show results for only one trajectory in Fig.~\ref{fig_sample_path},
and panels (e) and (f) are results of averages for $200$ trajectories. 
The filled circle (Simple) corresponds to the results 
for the simple Euler scheme,
and the empty boxes (LL) are those for the local linearization method.
In (a) and (b), the bandwidth of the kernel is $W = 0.06476$,
and in (c), (d), (e) and (f), $W = 0.3$.
Solid curves in left panels [panels (a), (c), and (e)] 
and in right panels [panels (b), (d) and (f)] 
correspond to $f^\mathrm{true}(x)$ and $g^\mathrm{true}(x)$ in Eq.~\eqref{eq_original_sde_2}, respectively.
Error bars in panels (e) and (f) are the standard deviations.
}
\label{fig_results}
\end{center}
\end{figure*}

We apply the two algorithms in Sec.~IV
to the discretely observed data in Sec.~II.
In the numerical experiments, we use the Gaussian kernel.

We should firstly set the kernel bandwidth $W$.
It would be possible to select the kernel bandwidth $W$ using some criteria,
for example, using a cross-validation method.
However, here we set $W$ heuristically.
A simple choice for $W$ is the optimized bandwidth for the kernel density estimator
discussed in Sec.~III.
In Figs.~\ref{fig_results}(a) and (b),
we show the results obtained from the simple estimation (Algorithm 1)
and the estimation based on the local linearization (Algorithm 2),
using $W = 0.06476$.
Because the bandwidth is narrow and the number of data points, $N$, is small,
the estimation results have large fluctuations, as shown in Figs.~\ref{fig_results}(a) and (b).

In the estimation based on the local linearization,
the second-order approximation is used,
and then the bandwidth $W$ could be taken larger
than that of the kernel density estimator;
intuitively, it would be reasonable to select a bandwidth
three or five times as large as the optimized bandwidth for the kernel density estimator.
Larger bandwidth $W$ enables us to use a larger number of {\it effective} data points for the estimation.
Of course, if large $W$ is employed, the `second-order' approximation becomes invalid,
and hence the estimation results would be worse.
We here set $W = 0.3$ ad hoc.
Figures~\ref{fig_results}(c) and (d) are the estimated results.
As expected, the estimation results become smooth because 
a larger number of {\it effective} data points is available.
Due to the simple assumption for Algorithm 1,
the estimated results are not good;
the usage of the larger bandwidth $W$ gives a kind of averaging effects.
In contrast, the estimation based on the local linearization (Algorithm 2)
gives good estimates.

The above results are only for one time-series data in Fig.~\ref{fig_sample_path}.
Next, we generated $200$ time-series data with the same parameters,
and checked the validity of our proposed method.
Figures~\ref{fig_results}(e) and (f) show results
of averages of the drift and diffusion coefficients
for $200$ trajectories, respectively.
Here, we used $W = 0.3$ for the estimations.
The error bars in Figs.~\ref{fig_results}(e) and (f) are standard deviations of the results.
From the results, we see that 
the estimation method based on the local linearization (Algorithm 2) works well,
especially for the estimation of the drift coefficients.

Based on results on other numerical experiments,
we comment on the choice of the bandwidth $W$ as follows;
the choice of the bandwidth $W$ should be small,
but if the number of data points is not enough,
a slightly large bandwidth would be better.
One may intuitively estimate the bandwidth from the probability density $\hat{p}(x)$.
Although a method of the choice of the bandwidth is beyond the scope of the present paper,
the above choice would be enough in a practical sense.

\section{Concluding remarks}

In the present paper, we developed a nonparametric estimation scheme
for a stochastic differential equation from a discretely-observed 
time-series data.
In order to make the estimation scheme,
a concept based on a kernel density estimation was extended
and a kernelized likelihood function was derived.
The word, `kernelized', means that an observed point
is distributed with a bandwidth.
In addition, 
we employ a local linearization method
in order to deal with the discrete property or sparsity
of the observations.
A `second-order' approximation was introduced,
in which the diffusion coefficient is restricted 
to be positive naturally.
This avoids adding any constraint for the maximization or minimization 
for the log-likelihood function in the algorithm.
Using a toy model, we demonstrated that the estimation method
based on the local linearization method works well.

Although results are not shown, we applied the estimation schemes to several different models,
and confirmed that they work well.
In addition, we performed numerical experiments 
for cases with larger $\Delta t_i$.
If we have larger $\Delta t_i$, results become worse.
This is because that the approximation of the local linearization is inadequate 
for the larger time interval.
For the large interval cases,
the Kramers-Moyal coefficients estimation would be available \cite{Honisch2011},
as explained in Sec.~I.
However, as stated before,
some additional computational costs are needed.
On the other hand,
our estimation scheme is based on an approximated analytical solution 
of a stochastic differential equation,
and then the computational time is largely reduced.
For example, a computational time for one point $x_0$ in Algorithm 2 
is a few seconds in a laptop computer.
In this sense, our estimation method is a complementary one with previous works.


Finally, we comment that
the estimation scheme for stochastic differential equations 
could be extended to multivariate cases straightforwardly,
because the local linearization method has already been formulated 
for multivariate cases \cite{Biscay1996}.
In addition, estimations in the presence of strong measurement noise
\cite{Bottcher2006,Kleinhans2007a,Lehle2011} should be considered in future works.


\section*{ACKNOWLEDGMENTS}

The author thanks T.J. Kobayashi for motivating this work.
This work was supported in part by grant-in-aid for scientific research 
(Nos. 20115009 and 21740283)
from the Ministry of Education, Culture, Sports, Science and Technology (MEXT), Japan.

\appendix

\section{Local Linearization method}

The local linearization method \cite{Biscay1996,Shoji1998}
is one of the useful approximations for nonlinear stochastic differential equations.
The basic concept of the local linearization method
is that the nonlinear stochastic differential equation
is approximated locally as a linear stochastic differential equation.
It would be possible to obtain the same consequence 
using a Fokker-Planck equation.
In Refs.~\cite{Biscay1996,Shoji1998},
an approximation for the stochastic differential equation is explicitly given,
and for the reader's convenience, we briefly review the local linearization method.
For details, see Refs.~\cite{Biscay1996,Shoji1998}.

For simplicity, we here consider a one-dimensional stochastic process $x_t$
satisfying 
\begin{align}
dx_t = A(x_t) dt + B(x_t) d B_t,
\end{align}
where $A(x_t)$ is twice continuously differentiable with respect to $x_t$,
$B(x_t)$ is a continuously differentiable function of $x_t$,
and $B_t$ is a standard Brownian motion.
The above stochastic differential equation can be transformed 
into a more tractable equation as follows:
\begin{align}
d z_t = \left( A \frac{d\phi}{dx} + \frac{B^2}{2} \frac{d^2 \phi}{dx^2} \right) dt + dB_t,
\label{eq_app_converted}
\end{align}
where $z_t = \phi(x_t)$
and $\phi(x_t)$ satisfies an ordinary differential equation $B \frac{d \phi}{dx} = 1$.
Ito's formula immediately gives Eq.~\eqref{eq_app_converted}.
Hence, we here only consider the following stochastic differential equation with 
a constant diffusion coefficient:
\begin{align}
d x_t = A(x_t)dt + d B_t.
\end{align}

In the local linearization method,
the drift term $A(x_t)$ is locally approximated by a linear function of $x_t$.
Using Ito's formula, we have
\begin{align}
dA =  \frac{1}{2} \frac{\partial^2 A}{\partial x^2}  dt
+ \frac{\partial A}{\partial x} dx.
\label{eq_app_dA}
\end{align}
Here, an assumption that both coefficients in Eq.~\eqref{eq_app_dA} are constant 
for a small interval $[s,t)$
gives
\begin{align}
&A(x_t) - A(x_s)   \nonumber \\
&= \frac{1}{2} \left. \frac{\partial^2 A}{\partial x^2} \right|_{x=x_s} (t-s) 
+ \left. \frac{\partial A}{\partial x} \right|_{x=x_s}(x_t - x_s).
\end{align}
Hence, we obtain the drift coefficient $A(x_t)$ as
\begin{align}
A(x_t) = L_s x_t + M_s t + N_s,
\end{align}
where
\begin{align*}
L_s =& \left. \frac{\partial A}{\partial x} \right|_{x=x_s}, \quad
M_s = \frac{1}{2} \left. \frac{\partial^2 A}{\partial x^2} \right|_{x_s}, \\
N_s =& A(x_s,s) - \left. \frac{\partial A}{\partial x}\right|_{x=x_s} x_s
- \left. \frac{1}{2} \frac{\partial^2 A}{\partial x^2} \right|_{x=x_s} s.
\end{align*}
Finally, we obtain a linear stochastic differential equation as follows:
\begin{align}
d x_t = (L_s x_t + M_s t + N_s)dt + d B_t.
\end{align}
The linear stochastic differential equation can be solved analytically,
and the solution is 
\begin{align}
x_t =& x_s + \frac{A(x_s)}{L_s} \left( e^{L_s(t-s)} -1\right) \nonumber \\
&+ \frac{M_s}{L_s^2} \left( e^{L_s(t-s)} -1 - L_s(t-s)\right)
+ \int_s^t e^{L_s (t-u)} dB_u,
\end{align}
where the fourth term follows the Gaussian distribution with mean $0$ and variance
$( \exp(2L_s(t-s)) - 1)/(2L_s)$.
As a result, $x_t - x_s$ follows
the Gaussian distribution with mean 
\begin{align*}
\frac{A(x_s)}{L_s} \left( e^{L_s(t-s)} -1\right)
+ \frac{M_s}{L_s^2} \left( e^{L_s(t-s)} -1 - L_s(t-s)\right)
\end{align*}
and variance $( \exp(2L_s(t-s)) - 1)/(2L_s)$.

Combining the above results, the variable transformation $z_t = \phi(x_t)$, and its Jacobian,
we finally obtain the conditional probability used in Eq.~\eqref{eq_LL_log_likelihood}

\end{document}